\documentclass[preprint,12pt,aps]{revtex4}
\usepackage{amsmath}

\input{tcilatex}

\begin{document}

\title{Error Function Attack of chaos synchronization based encryption
schemes}
\author{Xingang Wang, Meng Zhan, C.-H. Lai$^{\ast }$}
\affiliation{Temasek Laboratories, National University of Singapore, 10 Kent Ridge
Crescent, Singapore 119260\\
$^{\ast }$Department of Physics, National University of Singapore, Singapore
117542}
\author{Hu Gang}
\affiliation{Department of Physics, Beijing Normal University, Beijing 100875, China}

\begin{abstract}
Different chaos synchronization based encryption schemes are reviewed and
compared from the practical point of view. As an efficient cryptanalysis
tool for chaos encryption, a proposal based on the Error Function Attack is
presented systematically and used to evaluate system security. We define a
quantitative measure (Quality Factor) of the effective applicability of a
chaos encryption scheme, which takes into account the security, the
encryption speed, and the robustness against channel noise. A comparison is
made of several encryption schemes and it is found that a scheme based on
one-way coupled chaotic map lattices performs outstandingly well, as judged
from Quality Factor. \newline
\end{abstract}

\maketitle

\section{Introduction}

Chaotic systems and conventional cryptosystems share many common properties,
such as the sensitivity to initial conditions and parameters, high
complexity and unpredictability, stochastic and random-like behaviors, and
so on \cite{kocarev-ieee}. Shanon, in his seminal paper of cryptography,
wrote \cite{shanon}: "In a good mixing transformation ... functions are
complicated, involving all variables in a sensitive way. A small variation
of any one (variable) changes (the outputs)\ considerably." These
requirements, which are also known as the \textit{diffusion} and \textit{%
confusion} properties in conventional cryptography theory, are just some of
the fundamental characteristics of chaotic dynamical systems widely studied
in the last several decades. It is indeed a celebrated property of chaotic
systems that they can produce highly complex signals while having relatively
simple system structures. Whereas the conventional cryptography is mainly
based on some particular algebraic or number-theoretic operations,
chaos-based cryptography depends entirely on some physical laws, this makes
chaos cryptography not only easy to formulate and analyze in theory, but
also simply to design and operate in applications. Because of these
advantages, chaos encryption has been an extremely popular topic of
investigation, pursued in recent years by researchers from different fields.

The conventional cryptography works on discrete-value systems while chaotic
cryptography works on continuous-value systems. Because of this, the early
attempts in this field mainly focused on using chaotic systems as
pseudo-random number (PN) generators in discrete-value implementations \cite%
{matthews}. Later, different encryption algorithms were developed based on
the different properties of chaos: algorithms that build upon discretization
and mapping, ergodicity, perturbation and control, targeting, alphabet units
portion etc., have been proposed for different purposes \cite{algorithm}.
Each algorithm has its special characteristics and can be well utilized
under certain restrictions. In 1990, Pecora and Carroll published the first
paper on chaos synchronization and assumed its application in secure
communication \cite{pecora}. From then on, chaos synchronization based
encryption (CSE) has been a hot topic both in theory and in applications 
\cite{cuomo,roy,kocarev-1}. Comparing with other algorithms, the CSE
algorithm is advanced in many aspects. Firstly, the well developed theory of
chaos synchronization provides a solid basis for its feasibility and
performance analysis. Secondly, experiments show that chaos synchronization
can be robust and easily realizable in applications. Finally, it is easy to
synchronize high-dimensional chaotic systems constructed by coupling or
cascading low-dimensional systems, which has been used as a standard method
for generating highly complex signals in practical operations. As a result,
different encryption schemes based on the idea of CSE, such as chaos
masking, chaos key shifting and chaotic modulation etc., have been proposed
in recent years.

Initially, chaos encryption was implemented on low-dimensional chaotic
systems both theoretically and experimentally. These encryption methods are
reasonably fast, simple to implement, and have been assumed secure. However,
various security flaws have been found, flaws that allow the private message
to be extracted without a knowledge of the secret keys, or of the underlying
dynamics \cite{zhoucs,cerdeira,short}. In order to overcome these drawbacks,
recent developments have focused on hyperchaos systems which possess many
positive Lyapunov exponents (LEs) and are more complex in dynamics. In
particular, spatiotemporal chaos has been investigated widely for its
excellent performance in correlations and spread spectrum multiple access
communication \cite{ocml-x}. However, while the early works on chaos
encryption were discussed in the community of applied cryptography, recent
works are almost exclusively considered inside the nonlinear systems
community. Even a casual evaluation of these algorithms shows a lot of
potential pitfalls and inherent drawbacks \cite{kocarev-ieee,Dachselt}. For
a good practical encryption algorithm, security is not the only requirement;
other properties such as encryption speed and error rate should be
considered as well. Unfortunately, so far in the field of chaos encryption,
most efforts are still directed at improving system security, with the other
properties receiving little attention. In general, when the dimension of a
chaotic system is increased, its encryption speed will suffer, and for a CSE
system, the synchronization time will increase too. Together these will
effect the system cost and the global performance in different manners.
Therefore, it is necessary to analyze different chaos encryption schemes by
taking all these crucial aspects into account.

In this paper, we will analyze different chaos synchronization based
encryption schemes from the application point of view. For each scheme,
three properties: security, encryption speed, and error propagation length
(EPL), will be investigated and compared. We propose an efficient
cryptanalysis method, the Error Function Attack (EFA) method, for the
evaluation of chaos encryption systems. Moreover, we define a quantity,
which we call the Quality Factor (QF), that allows us to compare the
practical applicabilities of different cryptosystems. Based on the
simulation results, we give a rough ranking of various chaos encryption
schemes according to their feasibilities for application.

This paper will be arranged as follows. In Sec. II, we propose and analyze
the EFA method in detail. We then evaluate the security of different chaos
encryption schemes using the EFA method in Sec. III, and define the Quality
Factor. Comparison of the different encryption schemes using the QF is
presented in Sec. IV. A brief conclusion is then given in Sec. V.

\section{Error Function Attack}

A cryptosystem is usually designed in such a way that its security relies on
as few as possible secrets expressed as keys, and it is also believed that a 
\textit{public-structure} encryption system is more reliable than a secret
one \cite{bruce}. A fundamental assumption in cryptanalysis enunciated by
Kerckhoffs in the nineteenth century is that the security of an algorithm
must reside entirely on the key (the Kerckhoffs principle). According to
this assumption, for a good encryption algorithm, if Eve (the eavesdropper)
wants to extract the message transmitted between Alice (the transmitter) and
Bob (the receiver), the only thing she needs to do is try to find out the
secret key used; she will not need to spend time on extracting the system's
structure. In contrast, most chaos encryption schemes do not observe this
principle rigidly and the implementations try to hide everything from the
public. This approach has many negative impacts both on the system safety
and its commercial applications.

Based on the requirement of public-structure encryption, there are four
general types of cryptanalytic attacks, among them the \textit{%
known-plaintext attack} is the most common one. In the real world, it is
easy for an cryptanalyst to get some past plaintext messages that have been
encrypted. Hence, when we evaluate the securities of chaos encryption
schemes, vulnerability to the known-plaintext attack should be considered
first \cite{bruce}. In this paper, we will propose a known-plaintext and
public-structure cryptanalysis method, the Error Function Attack (EFA), for
the analysis of chaos encryption. Since the only thing Eve needs to do is to
find out the secret key $k$, the most straightforward approach would be for
her to try every possible key, $k^{^{\prime }}$, that Alice and Bob probably
can use (the keyspace). This is the so called \textit{brute-force} \textit{%
attack}. By defining the EFA function%
\begin{equation}
e(k^{^{\prime }})=\frac{1}{T}\int_{0}^{T}\left\vert D_{k^{^{\prime
}}}(C(t))-P(t)\right\vert dt=\frac{1}{T}\int_{0}^{T}\left\vert
D_{k^{^{\prime }}}(E_{k}(P(t)))-P(t)\right\vert dt,
\end{equation}%
Eve can scan the whole keyspace to find out the proper key $k^{^{\prime }}=k$
which satisfies $e(k^{^{\prime }})=0$. Here $T$ represents the amount of
data Eve will use, $dt$ represents the length of each block of plaintext ($%
P(t)$) and ciphertext ($C(t)$), $E_{k}(P)$ represents the encryption process
and $D_{k^{^{\prime }}}(C(t))$ denotes the decryption process.

For public-structure and known-plaintext attack, there are many well
developed methods in conventional cryptanalysis, such as the linear attack
and the differential attack. These methods have proved to be ineffective
when used on chaos encryption systems \cite{kocarev-1}. As we mentioned in
Sec. I, the basic difference between conventional cryptography and chaos
cryptography is that the conventional encryption is defined on discrete sets
and the chaos encryption is defined on continuous sets. This makes the
keyspace behavior of chaotic systems very different from that of
conventional systems \cite{kocarev-1,Fraser}. Due to its continuous-value
property, in chaos encryption systems keys that are not identical but are
very close can still be used to synchronize the two systems very well, thus
will form a \textit{key basin} around the actual secret key \cite%
{roy,zhoucs,Dachselt,Fraser}. Because of this, Eve will not need to try all
the keys in the keyspace; she can, according to the special characters of
key basin, find some optimization method to systematically adjust the trial
key $k^{^{\prime }}$ untill $k^{^{\prime }}=k$. By this way, the
cryptanalysis time can be greatly reduced.

In addition to this, EFA also attacks the weakest part in chaos encryption
systems -- the nonchaotic receiver. Since the dynamics of the receiver is
definitely not sensitive to the initial conditions and may not be sensitive
to the parameter mismatch as well, it is much easier to attack the receiver
than to attack the transmitter whose sensitivity depends on both parameters
and initial conditions. Thus, as a public-structure and known-plaintext
attack, EFA not only fully exploits the information of the known structure
and known plaintext, but also enjoys the advantage of considerable keyspace
reduction by attacking the nonchaotic receiver.

\section{EFA analysis on different chaos encryption schemes}

For any encryption scheme, the most important evaluation is about its
security. In Sec. II, we had mentioned that EFA has many advantages for
cryptanalysis of chaos based encryption. In this section, we will employ
this method to evaluate the security of some popular encryption schemes.
Specifically, the active-passive decomposition model, the piecewise linear
function model, the time delay model, the noise driving sequential
synchronization model, and the one-way coupled map lattices model. All these
models apply chaos synchronization and make use of high dimensional
hyperchaos, i.e. they are assumed to have better security than low
dimensional chaos. According to the requirement of the Kerckhoffs principle,
we consider only systems with the public-structure, i.e., the entire
dynamics is opened to the public except a single parameter which serves as
the secret key. Meanwhile, we assume that the eavesdropper could obtain some
past plaintext-ciphertext pairs, namely, we consider the public-structure
and known-plaintext attack. EFA will be used for analyzing all these
cryptosystems.

\subsection{Active-passive decomposition\ (APD) model}

Proposed in 1995 by Kocarev \textit{et al } and later generalized widely,
the active-passive decomposition (APD)\ model has been one of the most
popular chaos encryption schemes studied during the past years \cite%
{kocarev-2}. This model has two important characteristics: (a) the message
is not added directly to the chaotic carrier but drives the dynamical system
constituting the transmitter; (b) the dimension of encryption system can be
arbitrarily high by cascading identical or different individual systems, and
thus the APD\ model had been assumed with high security in early time. For
the consideration of the public-structure and known-plaintext attack, each
parameter in the equations can be regarded as a secret key in the APD model.
Here we use cascaded R\"ossler systems investigated in Ref. \cite{kocarev-2}
as our model:

Transmitter:%
\begin{eqnarray}
\overset{.}{x}_{1} &=&2+x_{1}(y_{1}-4),  \notag \\
\overset{.}{y}_{1} &=&-x_{1}-z_{1}, \\
\overset{.}{z}_{1} &=&y_{1}+a_{1}s_{1},  \notag
\end{eqnarray}%
with 
\begin{equation}
s_{1}=z_{1}+0.25P,
\end{equation}%
and 
\begin{eqnarray}
\overset{.}{x}_{i} &=&2+x_{i}(y_{i}-4),  \notag \\
\overset{.}{y}_{i} &=&-x_{i}-z_{i}, \\
\overset{.}{z}_{i} &=&y_{i}+a_{i}s_{i},  \notag
\end{eqnarray}%
with 
\begin{equation}
s_{i}=z_{i}+0.25s_{i-1}\text{, \ \ \ }i=2,3,....n.
\end{equation}%
Here $P$ stands for the message to be transmitted privately and $s_{n}$
represents the output ciphertext. All the parameters $a_{i}$, $i=1,2,...,n$,
can be used as secret keys and are chosen carefully to keep the last
oscillator staying within chaos. In order to simplify the related analysis,
we set $P(t)=0$ in Eq. (1) for each scheme investigated in this paper
without losing generality. In fact, the APD model performs a multiple
encryption process: the plaintext is first encrypted to $s_{1}$ by the first
R\"{o}ssler oscillator, then $s_{1}$ is regarded as the new input plaintext
and encrypted to $s_{2}$ by the second R\"{o}ssler oscillator, and the same
process repeated down to the last R\"{o}ssler oscillator which generates $%
s_{n}$ as the last output ciphertext. On the receiver side, the whole system
is identical to the transmitter except the encryption process be reversed,

Receiver:

\begin{eqnarray}
\overset{.}{x}_{n}^{^{\prime }} &=&2+x_{n}^{^{\prime }}(y_{n}^{^{\prime
}}-4),  \notag \\
\overset{.}{y}_{n}^{^{\prime }} &=&-x_{n}^{^{\prime }}-z_{n}^{^{\prime }}, \\
\overset{.}{z}_{n}^{^{\prime }} &=&y_{n}^{^{\prime }}+a_{n}^{^{\prime
}}s_{n}^{^{\prime }},  \notag
\end{eqnarray}%
with 
\begin{equation}
s_{n}^{^{\prime }}=s_{n},
\end{equation}%
and 
\begin{eqnarray}
\overset{.}{x}_{i}^{^{\prime }} &=&2+x_{i}^{^{\prime }}(y_{i}^{^{\prime
}}-4),  \notag \\
\overset{.}{y}_{i}^{^{\prime }} &=&-x_{i}^{^{\prime }}-z_{i}^{^{\prime }}, \\
\overset{.}{z}_{i}^{^{\prime }} &=&y_{i}^{^{\prime }}+a_{i}^{^{\prime
}}s_{i}^{^{\prime }},  \notag
\end{eqnarray}%
with 
\begin{equation}
s_{i}^{\prime }=(s_{i+1}^{\prime }-z_{i+1}^{\prime })/0.25\text{, \ \ \ }%
i=n-1,n-2.....1.
\end{equation}%
By setting $a_{i}^{^{\prime }}=a_{i}$, the receiver can be synchronized with
the transmitter and the message can be recovered step by step using Eq. (9),
and the original message will be recovered at the last oscillator by using $%
P=(s_{1}^{\prime }-z_{1}^{\prime })/0.25$.

Now we use EFA\ to analyze the key behavior of this encryption model.
Setting $a_{j}=0.45$, $j=1,2,...n$, the transmitter is spatiotemporally
chaotic with $n$ positive LEs. In our simulations we choose the most
effective (For the APD\ model, systems are cascaded unidirectionally. The
more last the parameter is, the more sensitive the decryption process will
be, this is also verified by our numerical simulations.) parameter $a_{n}$
the parameter of the last R\"{o}ssler oscillator, as the secret key for
encryptions and define the keyspace by keys that keep the last R\"{o}ssler
oscillator in the chaotic regime. For Eve, she has all the knowledge about
the transmitter and the receiver except $a_{n}$, and she also knows the
range of\ the keyspace. Consequently, she can run the receiver, Eqs.
(6)-(9), by choosing the trial key $a_{n}^{\prime }$ randomly within the
keyspace, and calculate the corresponding error function results by using
Eq. (1). In Fig. 1(a), we plot the key basin for the $n=2$ APD\ model by
using EFA%
\begin{equation}
e(a_{2}^{^{\prime }})=\frac{1}{T}\int_{0}^{T}\left\vert z_{1}^{^{\prime
}}(a_{2}^{^{\prime }})-z_{1}(a_{2})\right\vert dt.
\end{equation}%
It is found that the whole keyspace $a_{2}^{^{\prime }}\in \lbrack 0.44,0.46]
$ forms a single key basin, two straight lines with reversed slopes dominate
the behavior of $e(a_{2}^{\prime })$. The secret key $a_{2}^{\prime
}=a_{2}=0.45$ is located at the minimum point $e(a_{2}^{\prime }=a_{2})=0$.
With this structure, Eve can find the correct key easily through some
adaptive adjustments. For example, Eve can first try arbitrarily two trial
keys in the keyspace, $a_{2}^{\prime }(1)$ and $a_{2}^{\prime }(2)$, and by
comparing the respective values of $e$, she will know which direction she
should be adjusting her next attempt. In our simulations, only $6$ to $8$
tests are needed to find the properly location of $a_{2}$. Furthermore, by
using the slopes, we can evaluate $a_{2}$ proximately only by two trial
keys. We refer to this kind of key basin as the \textit{triangle basin} and
the method for key searching mentioned above as the \textit{adaptive
adjustment method }(AAM). It is obvious that this model has no security
against EFA.

In order to investigate the relationship between the dimension and the
security in this type of scheme, we also plot the key basins for $n=3,4,5$
in Fig. 1(b). As the dimension of the system increases, the key basin
changes only in shape, but the structure of triangle basin still persists.
This kind of basin structure can also be found in other APD\ based models,
and in Fig. 2 we plot the related key basins for the model used in Ref. \cite%
{kocarev-3}. All parameters and dynamics of the systems are those of the
original paper and the parameter in the equation of the first variable is
chosen as the key. We find that no matter how high the system dimension is
(the dimension changes from $5$ to $101$ in our simulations), the triangle
basin remains. (We should mention that as the system dimension is increased,
we also see a longer transient time before the triangle basin becomes
manifest.) Therefore, increasing the system size (i.e., the dimension of
hyperchaos)\ does not lead to an increase in security for this encryption
scheme. This result is rather surprising, and this behavior should be
seriously considered whenever one hopes to reach high security by increasing
the dimension of chaotic system, or say, by applying spatiotemporal chaos.

\subsection{Coupled piecewise linear function (CPLF) model}

For a long time, the piecewise linear function model has been another
popular nonlinear model investigated extensively \cite%
{piecewise-1,piecewise-2,piecewise-3}. There exists a well-developed theory
of piecewise linear maps which generate uniformly distributed signals, and
it is known that piecewise linear maps share nice properties of invariant
measures, ergodicity and statistical independence \cite{Dachselt}. In Ref. 
\cite{piecewise-2}, the authors proposed an efficient encryption scheme
based on coupled piecewise linear maps, and they declared that such
cryptosystems not only enjoy high security, but also have an "immense
parameter space"\ for key choosing even in lower dimensional encryption
systems. Here we choose the five-dimensional system used there as our model
and analyze its security by using EFA. The dynamics of the transmitter and
the receiver can be written as:

Transmitter:%
\begin{eqnarray}
x_{5}(k-1)
&=&f_{1}[x_{3}(k-1)]+a_{51}x_{1}(k-1)+a_{52}x_{2}(k-1)+a_{53}x_{3}(k-1) 
\notag \\
&&+a_{54}x_{4}(k-1)+k_{7}x_{2}(k-1)\sin [k_{8}x_{3}(k-1)],  \notag \\
x_{1}(k) &=&f_{2}[x_{5}(k-1)]+c\cdot P(k),  \notag \\
x_{2}(k)
&=&f_{3}[x_{3}(k-1)+x_{4}(k-1)]+x_{1}(k)+a_{22}x_{2}(k-1)+a_{23}x_{3}(k-1) 
\notag \\
&&+a_{24}x_{4}(k-1)+k_{1}x_{2}(k-1)\sin [k_{2}x_{3}(k-1)], \\
x_{3}(k)
&=&f_{4}[x_{4}(k-1)]+a_{32}x_{2}(k-1)+a_{33}x_{3}(k-1)+a_{34}x_{4}(k-1) 
\notag \\
&&+k_{3}x_{3}(k-1)\sin [k_{4}x_{4}(k-1)],  \notag \\
x_{4}(k) &=&a_{42}x_{2}(k-1)+a_{44}x_{4}(k-1)+k_{5}x_{4}(k-1)\sin
[k_{6}x_{2}(k-1)],  \notag
\end{eqnarray}%
with the piecewise linear map 
\begin{equation}
f_{i}(x)=C_{0}+C_{1}x+\underset{j=1}{\overset{N}{\sum }}D_{j}\cdot
\left\vert x-E_{j}\right\vert \text{, \ mod }\chi .
\end{equation}

Receiver:%
\begin{eqnarray}
x_{4}^{\prime }(k) &=&a_{42}^{\prime }x_{2}(k-1)+a_{44}^{\prime
}x_{4}^{\prime }(k-1)+k_{5}^{\prime }x_{4}^{\prime }(k-1)\sin [k_{6}^{\prime
}x_{2}(k-1)],  \notag \\
x_{3}^{\prime }(k) &=&f_{4}^{\prime }[x_{4}^{\prime }(k-1)]+a_{32}^{\prime
}x_{2}(k-1)+a_{33}^{\prime }x_{3}^{\prime }(k-1)+a_{34}^{\prime
}x_{4}^{\prime }(k-1)  \notag \\
&&+k_{3}^{\prime }x_{3}^{\prime }(k-1)\sin [k_{4}^{\prime }x_{4}^{\prime
}(k-1)],  \notag \\
x_{1}^{\prime }(k) &=&x_{2}(k)-f_{3}^{\prime }[x_{3}^{\prime
}(k-1)+x_{4}^{\prime }(k-1)]-a_{22}^{\prime }x_{2}(k-1)  \notag \\
&&-a_{23}^{\prime }x_{3}^{\prime }(k-1)-a_{24}^{\prime }x_{4}^{\prime
}(k-1)-k_{1}^{\prime }x_{2}(k-1)\sin [k_{2}^{\prime }x_{3}^{\prime }(k-1)],
\\
x_{5}^{\prime }(k-1) &=&f_{1}^{\prime }[x_{3}^{\prime }(k-1)]+a_{51}^{\prime
}x_{1}^{\prime }(k-1)+a_{52}^{\prime }x_{2}(k-1)  \notag \\
&&+a_{53}^{\prime }x_{3}^{\prime }(k-1)+a_{54}^{\prime }x_{4}^{\prime
}(k-1)+k_{7}^{\prime }x_{2}(k-1)\sin [k_{8}^{\prime }x_{3}^{\prime }(k-1)], 
\notag \\
P^{\prime }(k) &=&\{x_{1}^{\prime }-f_{2}^{\prime }[x_{5}^{\prime
}(k-1)]\}/c.  \notag
\end{eqnarray}

This is a delicately designed cascaded system where the plaintext $P$ is
added to the first variable and influences all the other variables through
couplings. The variable $x_{2}$ serves as the ciphertext and also acts as
the driver signal of the receiver for chaos synchronization. In the
piecewise linear map $f_{i}(x)$, $E_{1}<E_{2}<...E_{N}$ are breaking points
and $C_{1}=\frac{1}{2}(m_{0}+m_{N})$, $D_{j}=\frac{1}{2}(m_{j}-m_{j-1})$, $%
C_{0}$ are parameters used for model adjustment, and $m_{j}$ is the slope
for the $j$th segment. In our simulations, we set $N=20$, $\chi =100$ and
set $E_{j}$ uniformly distributed in $[0,\chi ]$ so as to simplifying our
analysis. We also set the slope $m_{j}=N\times \frac{\sqrt{5}-1}{2}>1$ so
that the system stays within the chaotic regime. With $C_{0}=-50$, the
function of one single piecewise linear map is plotted in Fig. 3(a).

In this cryptosystem, all coupling parameters and parameters in each map
function can be used as secret keys. Here we use the most sensitive
parameter $a_{44}$ as the secret key and keep all the other parameters
public. According to the requirement of convergence we fix the whole
parameter set as%
\begin{eqnarray}
a_{22} &=&0.1,\text{ }a_{23}=-0.09,\text{ }a_{24}=0.1,\text{ }a_{32}=0.1,%
\text{ }a_{33}=0.1,\text{ }a_{34}=0.2,  \notag \\
a_{42} &=&-0.23,\text{ }a_{51}=0.1,\text{ }a_{52}=0.3,\text{ }a_{53}=0.1,%
\text{ }a_{54}=0.31,  \notag \\
c_{1} &=&0.06,\text{ }c_{2}=3.4,\text{ }c_{3}=0,\text{ }c_{4}=12.9,\text{ }%
c_{5}=0.378,\text{ }c_{6}=-0.99, \\
c_{7} &=&0.001,\text{ }c_{8}=1.  \notag
\end{eqnarray}%
For $a_{44}=-0.89$, we plot in Fig. 3(b) the key basin for $a_{44}^{\prime }$
by using the EFA 
\begin{equation}
e(k^{\prime })=\frac{1}{T}\overset{T}{\underset{k=1}{\sum }}\left\vert
P(k)-P^{\prime }(k^{\prime })\right\vert .
\end{equation}%
It is found that there exists a key basin $a_{44}^{\prime }\in \lbrack
-0.895,-0.885]$ which occupies a parameter interval of order $10^{-2}$
around the actual secret key. Compared with the entire keyspace $%
a_{44}^{\prime }\in \lbrack -1,-0.75]$,\ which is of the order of $10^{-1}$,
it is still an relatively easy job to expose the secret key. For example, in
order to attack this system, Eve can divide the whole keyspace into some key
basin sized intervals, then test a few trial keys in each intervals to find
where the actual key basin is located. Once this interval is identified, Eve
can the focus her searching in this basin and determine the secret key by
AAM as used for the APD model. For the above mentioned five-dimensional
model, we can extract the secret key within $200$ tests, a considerably
reduced searching time in comparison to the brute-force attack which needs
about $10^{15}$ tests \cite{piecewise-2}.

Although still not quite secure enough, the security of the CPLF\ model has
been improved greatly compared to the APD model. Moreover, the security of
CPLF can be considerably improved by increasing the number of modulo
operations. With all other parameters unchanged, we plot in Fig. 3(c)\ and
(d)\ the EFA results for $C_{0}=0$ (the function of one single piecewise
linear map shown in Fig. 3(a) is divided into two separate parts this time)
and $m_{j}=10\times N\times \frac{\sqrt{5}-1}{2}$, respectively -- both
settings increase the number of modulo operations. It is apparent that the
system security is significant improved.

\subsection{Delay-differential equations (DDE)}

The time delay system employing delay-differential equations is an efficient
model for constructing high dimensional hyperchaos. Its dynamics structure
is rather simple, but its sequences can be very complex \cite{dde-1}.
Besides, DDE can be easily implemented in electronic systems \cite{dde-2}.
Comparing with other hyperchaotic systems, DDE has some special advantages
for encryption. First of all, DDE is an infinite dimensional dynamics with a
high dimensional attractor with many positive LEs; these guarantee high
complexity of its output signals. Secondly, the complexity and the number of
positive LEs can be controlled easily by adjusting the delay time $\tau $.
In these systems, we do not need to worry about the problem of weak keys,
since the system will always be chaotic and the number of positive LEs will
increase linearly, once $\tau $ exceeds the critical value. Finally,
synchronization between DDE has been established and proved to be robust 
\cite{dde-syn}. Therefore, DDE has been a popular model for encryption in
both theoretical and experimental investigations \cite{zhoucs,dde-3,dde-4}.
\ 

For this kind of cryptosystems, the delay time $\tau$ is the most suitable
choice for the secret key. Here we employ the Mackey-Glass DDE (MG DDE)
cryptosystem used in Ref. \cite{dde-5} as our model and analyze its security
by EFA. The dynamics of the transmitter and the receiver are

Transmitter:%
\begin{equation}
\frac{dx(t)}{dt}=-bx(t)+\frac{ax(t-\tau )}{1+x^{c}(t-\tau )}+\epsilon P(t),
\end{equation}

receiver:%
\begin{equation}
\frac{dy(t)}{dt}=-by(t)+\frac{ay(t-\tau ^{\prime })}{1+y^{c}(t-\tau ^{\prime
})}+[s(t)-y(t)],
\end{equation}%
with 
\begin{equation}
s(t)=x(t)+\epsilon P(t).
\end{equation}

It was shown that with parameters $b=0.1$, $a=0.2$, and $c=10$, the above
system will always be kept in the chaotic regime for $\tau >16.8$. For $\tau
=300$, there are altogether $15$ positive LEs and the system dimension is
about $30$. As $\tau$ increases, both the number of positive LEs and the
Kaplan-Yorke dimension increase \cite{dde-1}. In our simulations, we
consider the keyspace with the range $\tau ^{\prime }\in \lbrack 150,450]$
and choose $\tau =300$ as the actual secret key. The key basin is plotted in
Fig. 4 using the EFA%
\begin{equation}
e(\tau ^{\prime })=\frac{1}{T}\int_{0}^{T}\left\vert x_{\tau }(t)-y_{\tau
^{\prime }}(t)\right\vert dt,
\end{equation}%
In this case, the key basin is the range $\tau ^{\prime }\in \lbrack
230,340] $, and considering the O$(10^{2})$ size of the entire keyspace, it
is obviously an easy job to determine the secret key by AAM as in the cases
discussed above.

\subsection{Noise driven sequential synchronization (NDSS) model}

A hierarchically structured cryptosystem is proposed recently \cite{kim},
employing sequentially synchronized chaotic systems. Sequential
synchronization is attained by first feeding a noiselike signal to a
variable of the first transmitter and its receiver simultaneously and then
feeding a variable of the first transmitter and its receiver to a variable
of the second transmitter and its receiver, respectively, and repeating the
feedings of successive variables in sequence. Plaintext is added directly to
the variables to form the ciphertext on the transmitter side, and is
recovered by synchronization on the receiver side. This is different from
the encryption schemes mentioned above, as the plaintext here is not
involved in the dynamics. Such an encryption scheme appears to have high
security, which can be enforced selectively: different users can maintain
different security levels according to the synchronization level that can be
reached. Here we consider the cryptosystem composed of one Navier-Stokes
oscillator and one Lorenz oscillator, as used in Ref. \cite{kim}, with both
the transmitter and the receiver sharing the same dynamics,%
\begin{eqnarray}
\overset{.}{x} &=&-1.9x+4[\alpha _{1}y+\beta _{1}f(t)]z+4uv,  \notag \\
\overset{.}{y} &=&-7.2[\alpha _{1}y+\beta _{1}f(t)]+3.2xz,  \notag \\
\overset{.}{z} &=&-4.7z-7.0x[\alpha _{1}y+\beta _{1}f(t)]+k,  \notag \\
\overset{.}{u} &=&-5.3u-xv,  \notag \\
\overset{.}{v} &=&-v-3.0xu,\text{ \ \ \ \ (Navier-Stokes)} \\
\overset{.}{p} &=&\sigma \lbrack (\alpha _{2}q+\beta _{2}z)-p],  \notag \\
\overset{.}{q} &=&cp-(\alpha _{2}q+\beta _{2}z)-pr,  \notag \\
\overset{.}{r} &=&p(\alpha _{2}q+\beta _{2}z)-br,\text{ \ \ (Lorenz)}  \notag
\end{eqnarray}%
where $\alpha _{1}$, $\alpha _{2}$, $\beta _{1}$ and $\beta _{2}$ are the
couplings, $f(t)\ $is the noise signal which reads%
\begin{equation}
f(t)=50\xi \sin [2\pi \times 0.8(\frac{1}{2}+\xi ^{^{\prime }})t]
\end{equation}%
where $\xi $ and $\xi ^{^{\prime }}$ are pseudorandom numbers within $(0,1)$.

With the parameters $k$, $\sigma $, $c$, and $b$ taken to be $36$, $10.0$, $%
28.0$, and $8/3$, respectively, for $\alpha _{1}=1.2$, $\beta _{1}=0.9$, $%
\alpha_{2}=0.9$, and $\beta _{2}=22.5$, both the transmitter and the
receiver Exhibit chaotic behavior but can be synchronized. For
public-structure and known-plaintext attack, we choose the parameter $k=36$
in the Navier-Stokes equations as the secret key and consider the keyspace $%
k^{\prime}\in \lbrack 35,37]$, where the whole system stays in the chaotic
regime and synchronization between the transmitter and the receiver can be
achieved. In our simulations, we choose the variables $v$ and $r$ as the
carriers, and use the EFA 
\begin{equation}
e(k^{\prime })=\frac{1}{T}\int_{0}^{T}\left\vert v^{\prime }(k^{\prime
})-v(k)\right\vert dt,\text{ }
\end{equation}%
and%
\begin{equation}
e(k^{\prime })=\frac{1}{T}\int_{0}^{T}\left\vert r^{\prime }(k^{\prime
})-r(k)\right\vert dt.
\end{equation}%
The key basins are plotted for the Navier-Stokes system and the Lorenz
system in Fig. 5(a)\ and Fig. 5(b), respectively. Again we find the triangle
basin in the Navier-Stokes system, with a similar one for the Lorenz system,
except with a little distortion. (This appears to be typical for every
variable in this system chosen as the secret key.) The conclusion is clear:
the secret key can be easily determined using AAM just as in the cases that
we have discussed, and the claim for high security does not seem to be
justified. From the results of our simulations, we do not see any
improvement with more complicated coupled chaotic systems.

\subsection{One-way coupled map lattices (OCML)}

For a long time, coupled map lattices (CML)\ have been used to investigate
the complex behavior of spatiotemporal chaos in many fields of nonlinear
science \cite{ocml-fh}. Recently, this kind of system has been utilized for
secure communication in a number of encryption algorithms. In particular,
the one-way coupled map lattices (OCML)\ is extensively used for
self-synchronizing, spatiotemporal chaos-based cryptosystems \cite%
{ocml-hu-1,ocml-hu-2}.

The earlier works on OCML\ inherited the classical ideas of chaos
encryption: they regarded OCML\ as a special spatiotemporal chaos system
with inherent high computational complexity and yet amendable to easy
analysis \cite{ocml-hu-1}. Later, modified OCML\ models were proposed to
make the systems more feasible for encryption application. These
modifications include the adoption of self-synchronization, the use of
binary sequences more suitable for digital communications \cite{ocml-x}, and
the application of modulo operations to improve the system security \cite%
{ocml-hu-2}, etc. Further studies show that OCML\ ciphers can be improved to
have not only competitive security when compared against conventional
ciphers, but also respectable encryption speed and low bit error\cite%
{ocml-hu-2}. In the following, we shall mainly evaluate the property of
security in a OCML\ system by employing the EFA.

We use a modified two-dimensional OCML\ encryption system recently proposed
in Ref. \cite{ocml-hu-3} as our model. The transmitter is comprised of two
parts: the first part is a one-dimensional OCML with $m$ coupled lattices, 
\begin{eqnarray}
x_{0}(n) &=&S_{N/2,N/2}(n)/2^{v},  \notag \\
x_{l}(n+1) &=&(1-\varepsilon )f_{l}[x_{j}(n)]+\varepsilon
f_{l-1}[x_{l-1}(n)], \\
f_{l}(x) &=&a_{l}x(1-x)\text{, \ \ }l=1,2,...m,  \notag
\end{eqnarray}%
while the second part is a two-dimensional OCML\ driven by the first part
through lattice $z$,%
\begin{eqnarray}
z(n+1) &=&(1-\varepsilon )f[z(n)]+\varepsilon f_{m}[x_{m}(n)],  \notag \\
y_{0,0}(n) &=&z(n)\times 2^{h}\text{ \ \ mod 1,}  \notag \\
y_{1,0}(n+1) &=&(1-\varepsilon )f[y_{1,0}(n)]+\varepsilon f[y_{0,0}(n)], 
\notag \\
y_{0,1}(n+1) &=&(1-\varepsilon )f[y_{0,1}(n)]+\varepsilon f[y_{0,0}(n)], 
\notag \\
y_{i,0}(n+1) &=&(1-\varepsilon )f[y_{i,0}(n)]+\varepsilon
\{0.8f[y_{i-1,0}(n)]+0.2f[y_{i-2,0}(n)]\}\text{, }i=2,..N, \\
y_{0,j}(n+1) &=&(1-\varepsilon )f[y_{0,j}(n)]+\varepsilon
\{0.2f[y_{0,j-1}(n)]+0.8f[y_{0,j-2}(n)]\}\text{, }j=2,..N,  \notag \\
y_{i,j}(n+1) &=&(1-\varepsilon )f[y_{i,j}(n)]+\varepsilon
\{0.5f[y_{i-1,j}(n)]+0.5f[y_{i,j-1}(n)]\}\text{, }2\leq i+j\leq N,  \notag \\
f(x) &=&4x(1-x).  \notag
\end{eqnarray}%
The output ciphertext reads%
\begin{eqnarray}
S_{i,j}(n) &=&[K_{i,j}(n)+P_{i,j}(n)]\text{ \ \ \ \ }\func{mod}\text{ }2^{v},
\notag \\
K_{i,j}(n) &=&\{\text{int}[y_{i,j}(n)\times 2^{\eta }]\}\text{ \ \ }\func{mod%
}\text{ }2^{v}\text{, \ 2}\leq i+j\leq N.
\end{eqnarray}%
The dynamics of the receiver is identical to that of the transmitter. When
the two systems are synchronized, messages can be recovered by the reverse
process of Eqs. (26). The detailed explanation on the structure of Eqs. (25)
and Eqs. (26) is found in Ref. \cite{ocml-hu-3}, and in this work we shall
focus on analyzing its security and other encryption properties. For this
cryptosystem, the parameters $a_{l}$, $l=2,3,...m$, can be used as the
secret keys. In the following simulations, we set the parameters $N=6$, $%
m=3, $ $\varepsilon =0.99$, $h=26$, $\eta =60$, $v=30$, $a_{l}=3.9$, $%
l=2,3,...m$, and choose just one parameter (just as we did for the previous
four models), $a_{1}$, as the secret key. Except for the lattice $y_{0,0}$,
each lattice in the second part can be used as an encryption unit, thus
significantly enhancing the encryption speed. (With the above settings, for
example, the system can generate a total of $25$ sequences simultaneously.)
Moreover, the modulo operations used in Eqs. (25) and Eqs. (26) also serve
to greatly improve the system security. This is therefore a system that
potentially has both high security and fast encryption speed.

With the requirement of public-structure and known-plaintext, we study
system's key basin by using the EFA%
\begin{equation}
e_{i,j}(a_{1}^{\prime })=\frac{1}{T}\overset{T}{\underset{n=1}{\sum }}%
\left\vert K_{i,j,a_{1}}(n)-K_{i,j,a_{1}^{\prime }}^{\prime }(n)\right\vert ,
\end{equation}%
with $K_{i,j,a_{1}^{\prime }}^{\prime }$ representing the corresponding
output in the receiver for the trial key $a_{1}^{\prime }$. In Fig. 6 we
plot the simulation results for $e_{3,3}$ with the secret key chosen as $%
a_{1}=3.9$. The key basin is observed to be within the range $[3.9-4\times
10^{-12},3.9+4\times 10^{-12}]$. Since the whole keyspace for this kind of
system is at least within range $a_{1}^{\prime }\in \lbrack 3.6,4.0]$, it
needs at least $10^{11}$ tests to determine the correct secret key using
EFA. We have also investigated through simulations the key basins for other
output sequences $S_{i,j}$, and found that the widths of the key basins
typically range from $10^{-11}$ to $10^{-12}$, depending on where the output
lattice is located. This suggests that any one output sequence is a good
candidate for encryption. In comparison with other four models mentioned
above, this model appears to possess much higher system security.

\section{Quality Factor}

In conventional cryptography, three criteria are commonly considered when
designing an effective and applicable cryptosystem: security, the encryption
speed, and the error rate \cite{bruce}. For chaos based encryption,
relatively little attention has been paid to the latter two criteria \cite%
{kocarev-ieee,Dachselt, Fraser}, which are intimately linked to the attempts
to achieve greater security. First of all, security of a cryptosystem
depends mainly on its dimension: the higher the dimension, the more complex
the signal it can generate. However, high dimensional systems usually incur
greater encryption time cost in both software and hardware implementations.
Obviously there will have to be a trade-off between these two in considering
the realistic application of a system. Furthermore, higher system security
usually implies higher system sensitivity, and this brings about another
problem for chaos encryption, name the system stability. Both the encryption
systems and the communication channels can be disturbed unavoidably for
various reasons like intentional attacks or unintentional perturbations).
The system sensitivity of a high security system will amplify these
disturbances rapidly. This is particularly problematic for chaos
synchronization based cryptosystems which have to spend a certain time to
eliminate these disturbances before any correct communication can be
accomplished. All these highlight the relevance of considerations beyond
just system security in evaluating the overall performance and applicability
of any given encryption scheme. In this section, we will analyze the
encryption models used in Sec. III by studying their performance in all the
three aspects mentioned above, and compare their advantages and drawbacks
from the point of application.

Before going further, we need to give the security a quantitative
description. Based on the analysis of EFA, there always exists a key basin
around the actual secret key. Without the knowledge of this basin, one may
not be able to extract the secret key by brute-force or other means of
searching.\ We define the key basin width, denoted by $w$, by the distance
between two trial keys located on the two sides of the key basin, which
exceeds the average EFA result of the entire keyspace (see the marked
regions in Fig. 3(b), Fig.4, and Fig. 6). Here, we call the number of
intervals (which are potential key basins) with this width in the whole
keyspace as the key number. (For different choices of the secret key, there
can be small differences in the basin widths.) The security of the system is
broken as soon as the actual key basin is exposed, since the searching cost
of extracting the secret key then once is negligible. This motivates us to
define the security of a cryptosystem as the entropy of the key number \cite%
{Fraser}%
\begin{equation}
S=\log _{2}\frac{K}{w},
\end{equation}%
with $K$ denoting the length of the entire keyspace and $\frac{K}{w}$ being
the key number.

We define the amount (megabit) of data that can be encrypted in each second
as the \textit{encryption speed}, denote it by $V$. The number $L$ of bits
that are affected when one bit is in error in the ciphertext is defined as
the \textit{error propagation length} (EPL). Our simulation study is based
on software implementation, and treat each variable as a $8$-byte (i.e. $64$%
-bit, double precision, real) data. In computing the EPL, random
perturbation is added within the range of the output signals, and the
average over $1000$ runs is taken.

As we had emphasized above, there exist some trade-off relations between
security, encryption speed and EPL, so from the point of application, we
need to strike a balance between these aspects so as to obtain a better
overall performance. In this paper, we represent this balance by the \textit{%
\ Quality Factor}, which we define as%
\begin{equation}
B=\frac{VS}{L}.
\end{equation}%
Although this definition may not reflect the precise relationship in real
applications, the Quality Factor does parameterize the interplay of the
factors affecting the overall performance of a cryptosystem in a concrete
and reasonable way. but it can fairly well reflect some basic properties for
evaluating system's overall performance. We find the following results for
the different encryption models discussed in Sec. III:

\begin{center}
\begin{tabular}{|c|c|c|c|c|c|}
\hline
& APD & CPLF & DDE & NDSS & OCML \\ \hline
$S$ & $\log _{2}1$ & $\log _{2}10^{3}$ & $\log _{2}10^{1}$ & $\log _{2}1$ & $%
\log _{2}10^{11}$ \\ \hline
$V$ & $31.6$ & $2.8$ & $24.5$ & $22.6$ & $216.7$ \\ \hline
$L$ & $1737$ & $681$ & $719$ & $2213$ & $22$ \\ \hline
$B$ & $0$ & $0.041$ & $0.113$ & $0$ & $360$ \\ \hline
\end{tabular}%
\ \ \ \ 
\end{center}

The above results are obtained on SGI OCTANE workstation (two 195MHZ IP30
CPU 256M RAM, Fortran90 compiler). We have also carried out computations on
other workstations and compilers and obtained qualitatively similar results.
For the simulations, we choose the parameters as for Fig. 1(a)\ for the APD
model, Fig. 3(a) for the CPLF model, Fig. 4 for the DDE model, Figs. 5(a)\
and (b) for the NDSS model, and Fig. 6 for the OCML model, respectively, so
as to maximize the overall performance for each scheme.

Based on the above table, we wish to make the following observations.

\begin{enumerate}
\item From the security point of view, the OCML and CPLF\ models perform
much better than the other models. This is mainly due to the modulo
operations employed in these two schemes. The modulo operation disposes the
most significant digits in the signals and keep only those less significant
digits, making the systems more sensitive to the key values. As
verification, we have also tried applying modulo operations on the output
signals of the NDSS model (the same parameters as in Figs. 5(a) and (b)),
and the related EFA results are plotted in Fig. 7(a)\ and (b)\ for the
Navier-Stokes system and the Lorenz system, respectively. The improvement in
the system security is obvious.

\item The OCML\ model possesses excellent encryption speed due to its simple
functions and multiple outputs. In our simulation, we find that although the
modulo operation can improve appreciably the system security greatly, it is
also a time-consuming function. This is largely responsible for the low
encryption speed of the CPLF\ model. In contrast, the $25$ simultaneous
outputs and the simplicity of the mapping functions more than compensate for
the extra processing time linked to the modulo operations.

\item Finally, from the results on the EPL, it is clear that lower
dimensional, mapping systems perform better than high-dimensional,
differential systems. Moreover, systems with multiple outputs but with only
one as the driving signal has a clear advantage in EPL performance. While
the errors in the transmitted bits of the driving signal are responsible for
the EPL, the remaining nondriving ciphertext bits do not give rise to such a
problem, i.e., one bit error of the transmitted ciphertext causes only one
bit error in the received plaintext. Together with the two dimensional
structure design and strong couplings, these features give the OCML\ model
an outstanding performance as far as EFL is concerned.
\end{enumerate}

\section{Conclusion}

In this paper, we proposed an efficient cryptanalysis tool, the EFA method,
to study the security of some well-known chaos encryption schemes. Our
systematic comparison suggests that even models (such as the APD and NDSS
models) with high dimensionalities (and hence the supposed higher security)
fail to maintain their security under the EFA. We are also of the opinion
that in addition to security consideration, there are other important
aspects of cryptosystems which affect their overall Performance. More
specifically, we consider, additionally, the encryption speed and the error
propagation Length, and suggest in this paper a quantity (the Quality
Factor) as a possible measure of the overall performance, or applicability,
of chaos-based cryptosystems. Through comparisons, we find that the modified
OCML model has the best overall performance among the models considered, and
some the reasons responsible for this performance are briefly
discussed.

Figure Captions:

Fig. 1 (a) Key basin of $a_{2}^{\prime }$ for the APD\ model composed of two
coupled R\"{o}ssler oscillators, $a_{2}=0.45$ is chosen as the secret key.
(b) Key basin of $a_{3}^{\prime }$, $a_{4}^{\prime }$, and $a_{5}^{\prime }$
for APD\ models with $3$, $4$, and $5$ coupled R\"{o}ssler oscillators,
respectively.

Fig. 2 Key basins for the model used in Ref. \cite{kocarev-3}.

Fig. 3 The coupling $a_{44}$ in Eqs. (11) is chosen as the secret key. (a)\
Map of CPLF for $C_{0}=-50$ and $m_{j}=20\times \frac{\sqrt{5}-1}{2}$, $%
j=1,2,...,20$. (b)\ Key basin of CPLF model with the same parameters as for
(a). (c)\ Key basin for $C_{0}=0$, and, (d)\ key basin for $m_{j}=10\times
20\times \frac{\sqrt{5}-1}{2}$, $j=1,2,...,20$, other parameters being the
same as for (a).

Fig. 4 Key basin for the DDE\ model, with $b=0.1$, $a=0.2$, and $c=10$. For
The secret key $\tau =300$, there are $15$ positive LEs and system dimension
is about $30$.

Fig. 5 The secret key $k=36$ is chosen here, and the difference between the
trial key $k^{\prime }$ and the secret key $k$, $\Delta k=k^{\prime }-k$, is
used as variable for the horizontal axis. (a)\ The key basin for the
Navier-Stokes system, (b)The key basin for the Lorenz system.

Fig. 6 The key basin for the OCML\ model with the chosen secret key $%
a_{1}=3.9$. The variable for the horizontal axis is the difference between
the trial key $a_{1}^{\prime }$ and the actual secret key $a_{1}$, $\Delta
a=a_{1}^{\prime }-a_{1}$.

Fig. 7 The key basins after the introduction of modulo operations for (a)
the Navier-Stokes system and (b) the Lorenz system, respectively. The
parameters are the same as for Fig. 5

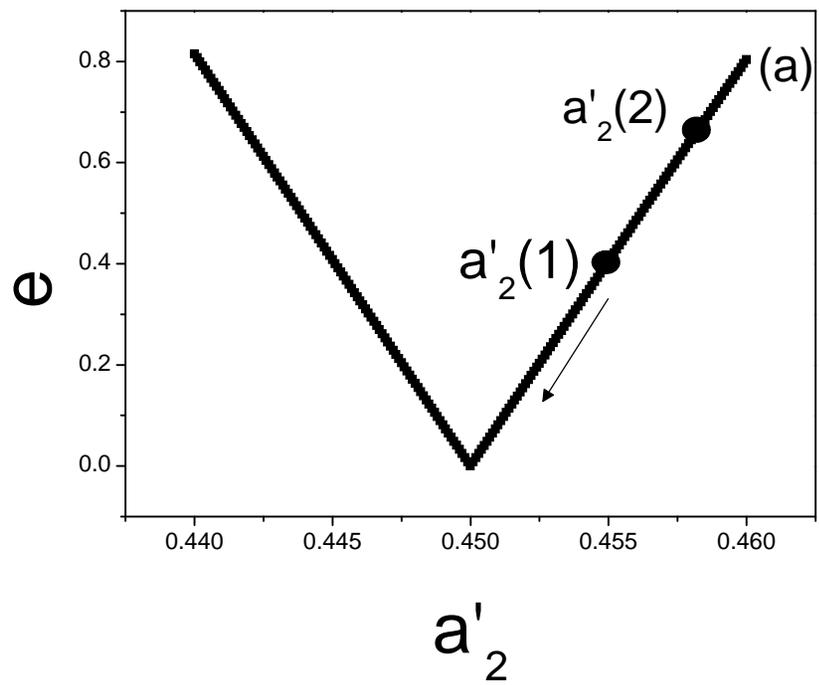
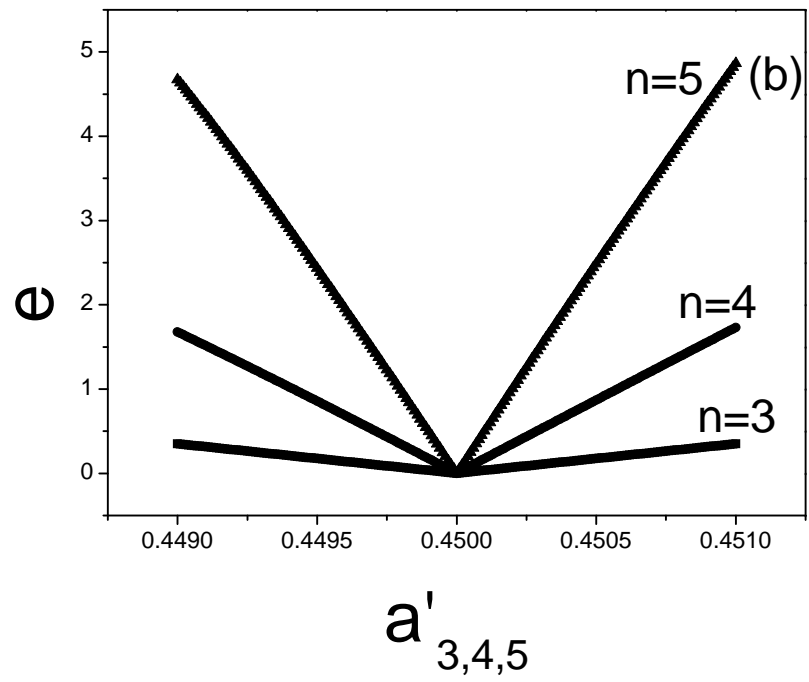
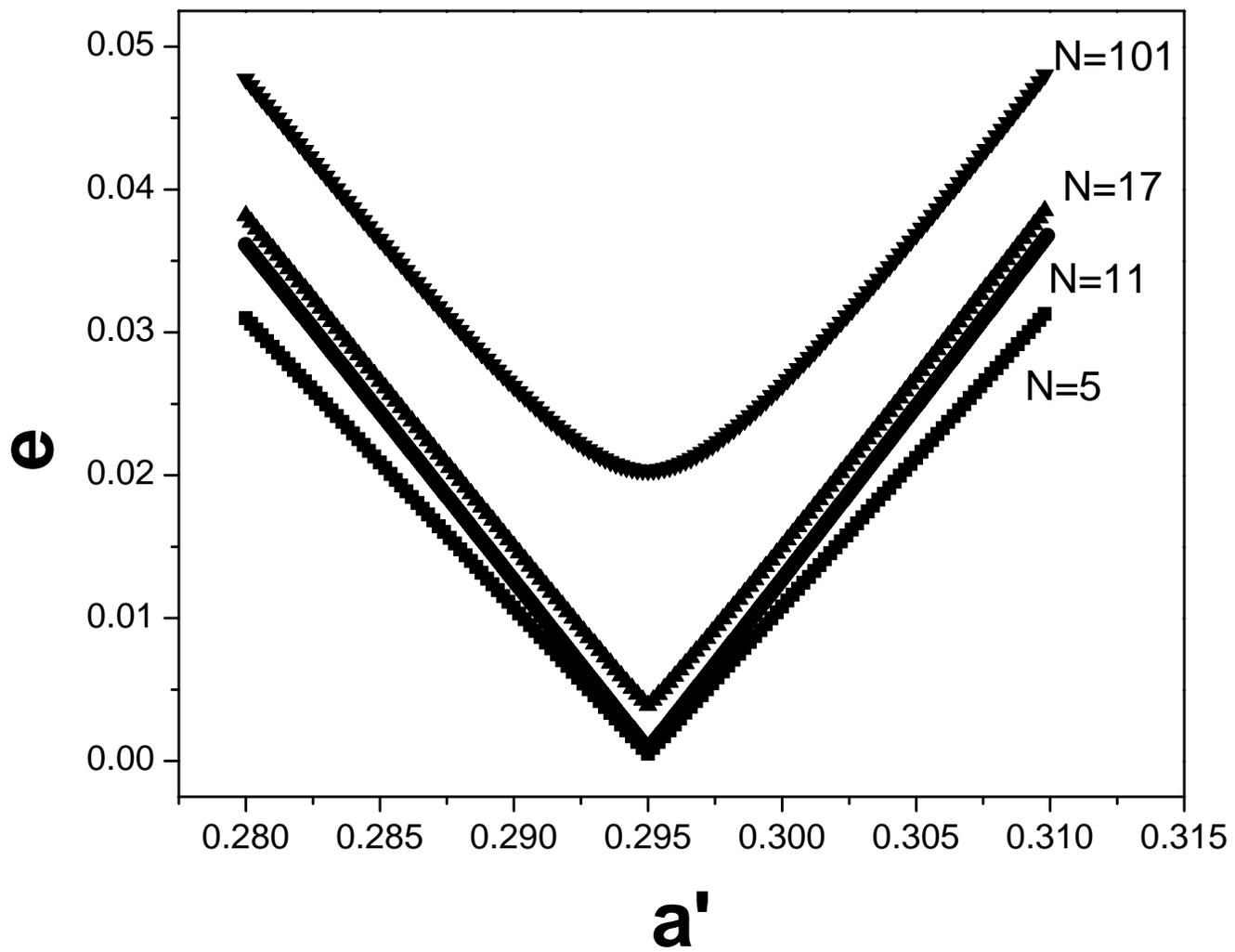
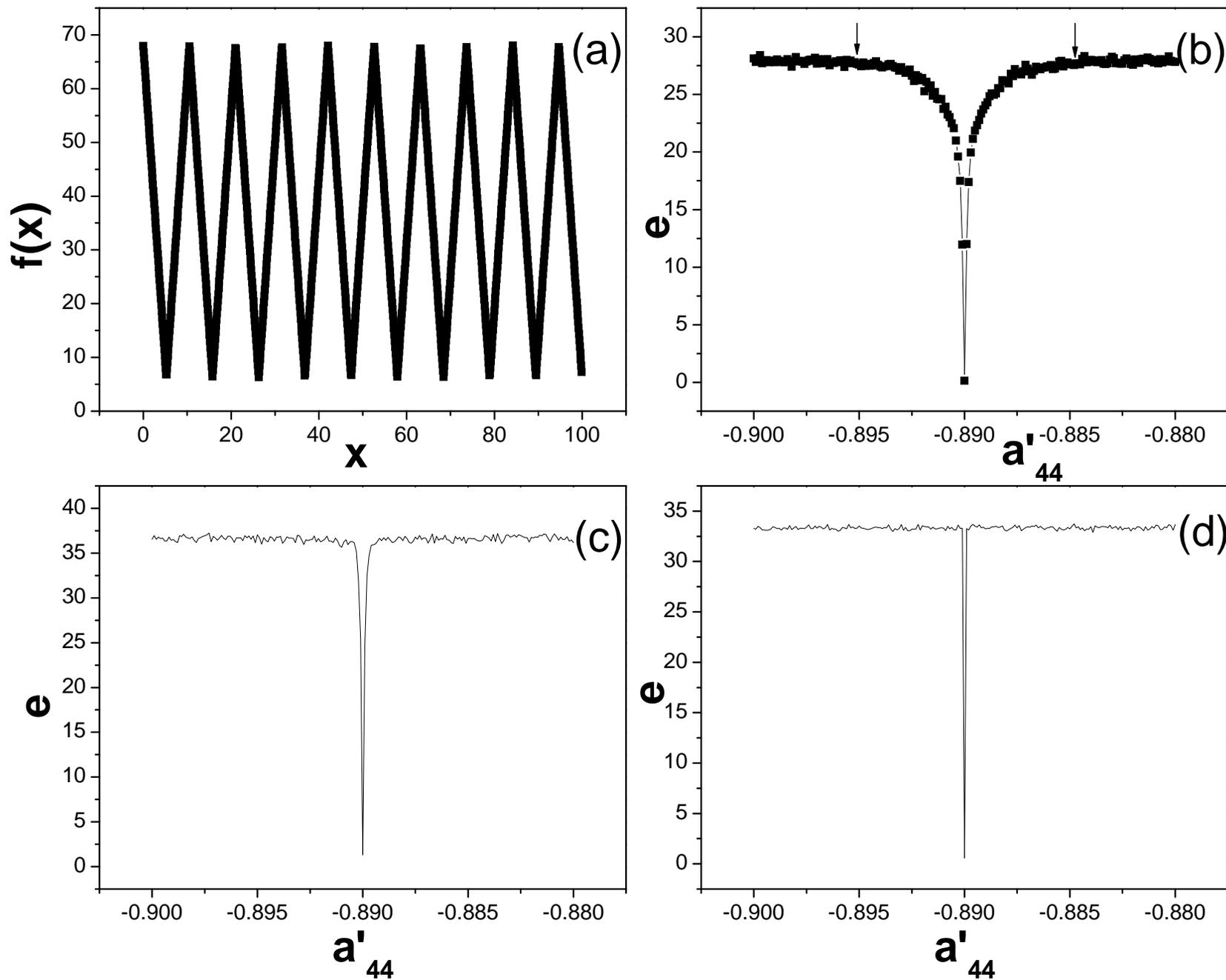
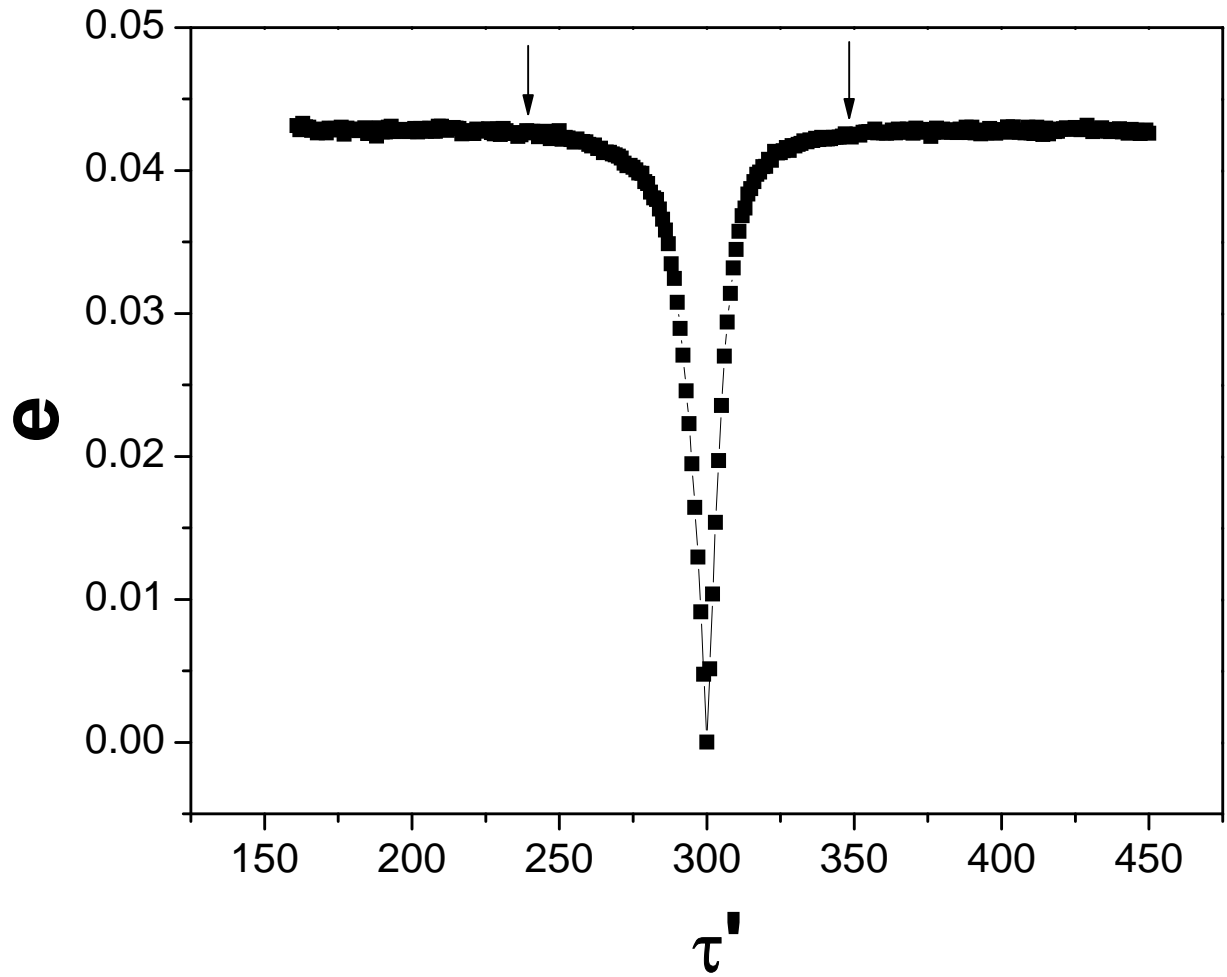
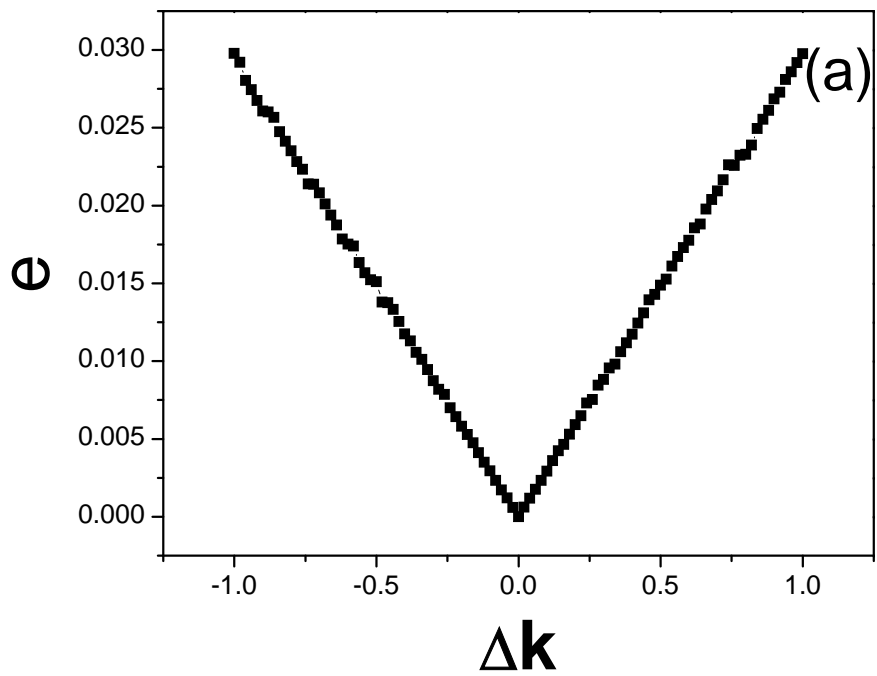
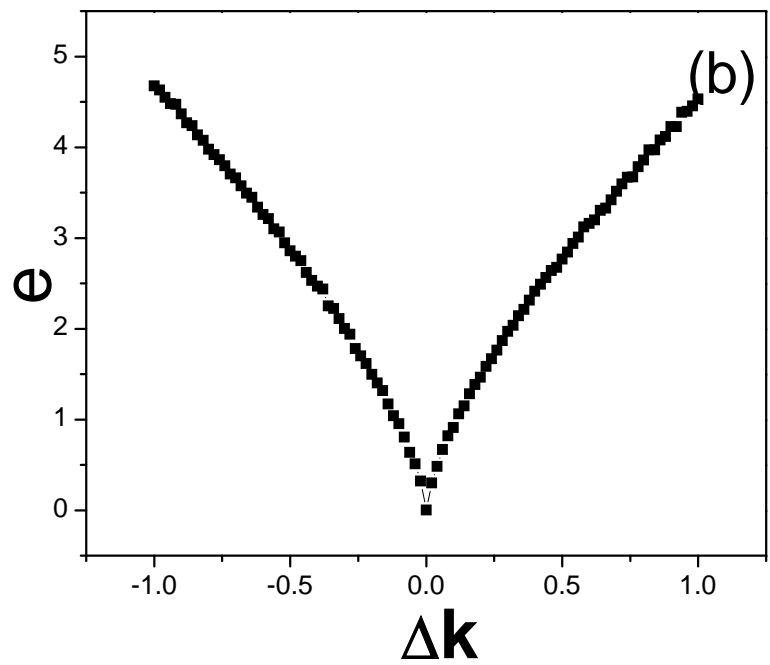
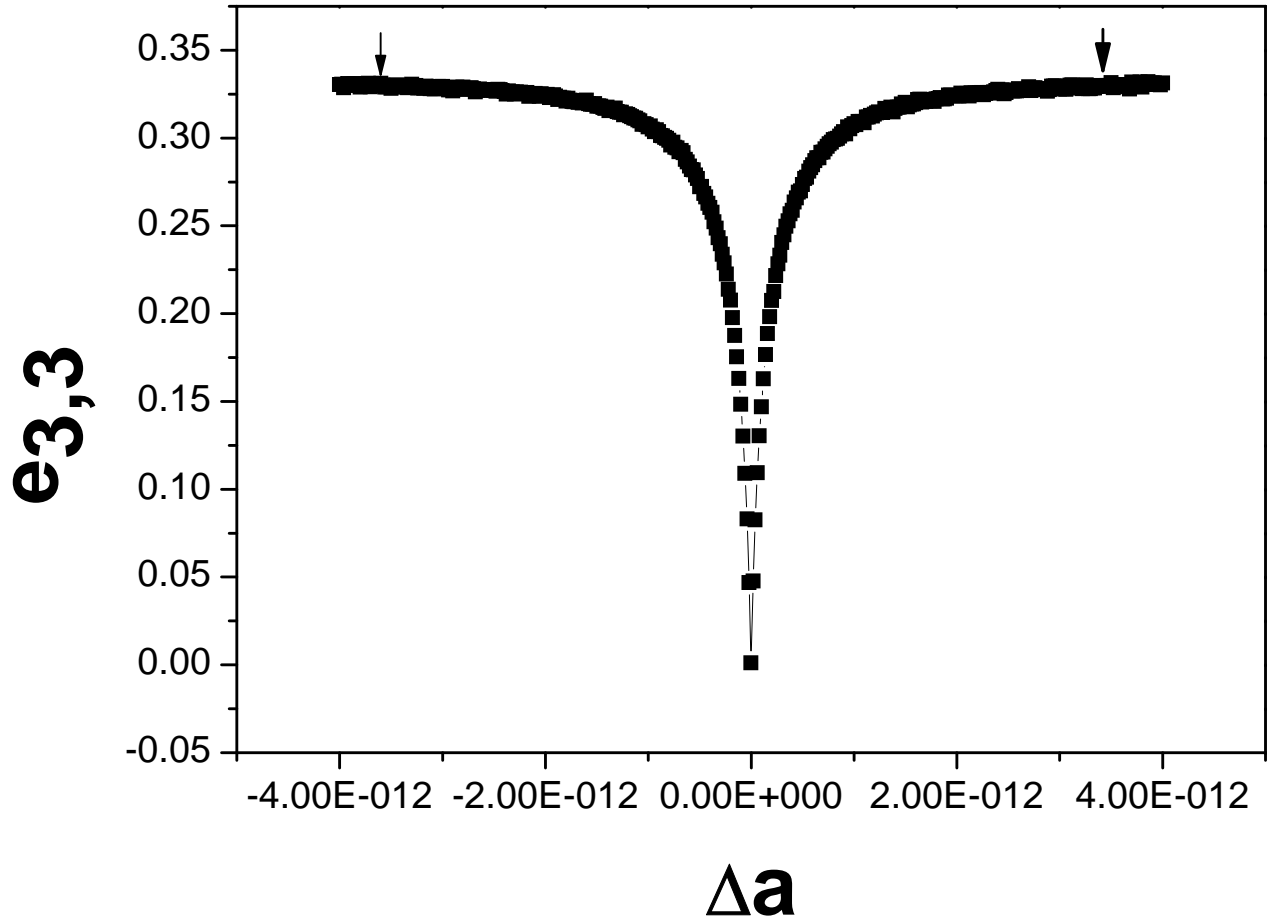
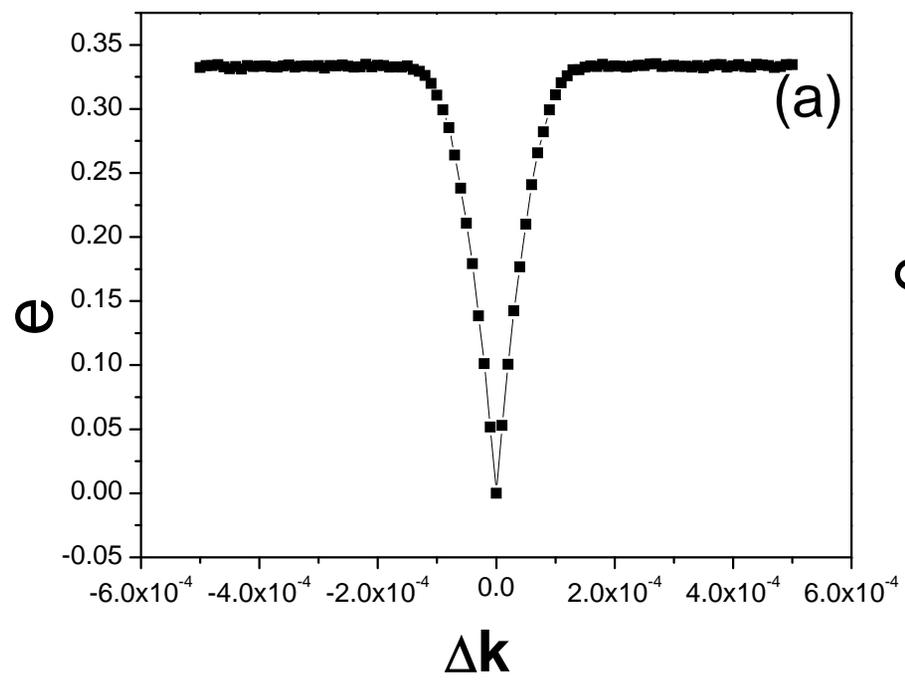
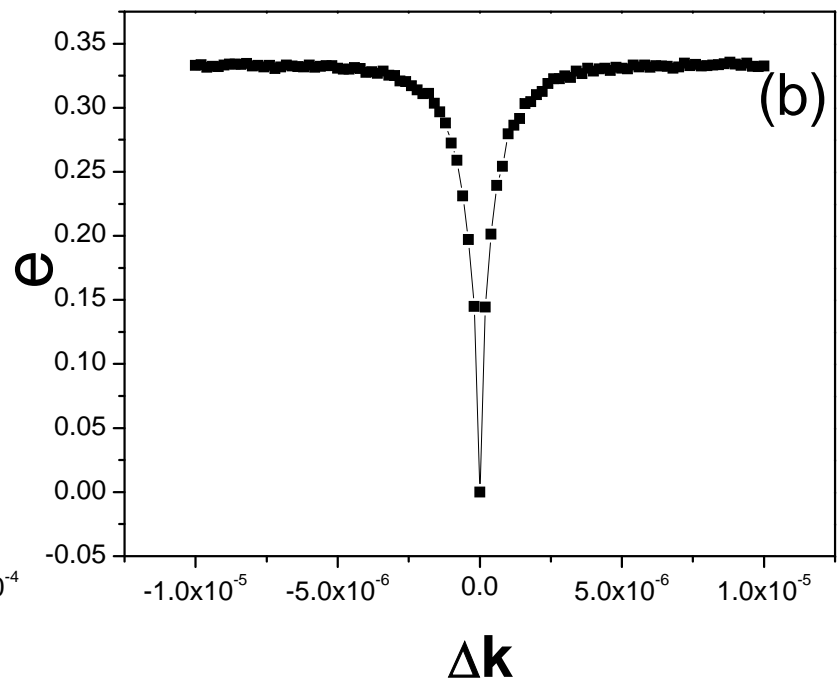
\end{document}